\documentclass{aip-cp}

\usepackage[numbers,compress]{natbib}
\usepackage{graphicx}
\begin{document}

% Title portion
\title{Instrumental Resolution of the Chopper Spectrometer 4SEASONS
Evaluated by Monte Carlo Simulation}

\author[aff1]{Ryoichi Kajimoto\corref{cor1}}
\author[aff2,aff3]{Kentaro Sato}
\author[aff1]{Yasuhiro Inamura}
\author[aff3]{Masaki Fujita}

\affil[aff1]{Materials and Life Science Division, J-PARC Center, Japan
Atomic Energy Agency, Tokai, Ibaraki 319-1195, Japan}
\affil[aff2]{Department of Physics, Tohoku University, Sendai 980-8578, Japan}
\affil[aff3]{Institute for Materials Research, Tohoku University
980-8577, Japan}
\corresp[cor1]{Corresponding author: ryoichi.kajimoto@j-parc.jp}

\maketitle

\begin{abstract}
We performed simulations of the resolution function of the 4SEASONS
spectrometer at J-PARC by using the Monte Carlo simulation package
McStas. The simulations showed reasonably good agreement with analytical
calculations of energy and momentum resolutions by using a simplified
description. We implemented new functionalities in Utsusemi, the
standard data analysis tool used in 4SEASONS, to enable visualization of
the simulated resolution function and predict its shape for specific
experimental configurations.
\end{abstract}

\section{INTRODUCTION}

Estimating the instrumental resolution function of neutron scattering
instruments is useful to determine physical quantities from experimental
data. For inelastic neutron scattering instruments, the resolution
function is expressed as an ellipsoid in the four-dimensional (4D)
momentum and energy space~\cite{cooper_acta}. For time-of-flight chopper
spectrometers, the analytical method for calculating instrumental
resolution has been established for a simple spectrometer
\cite{perring_icans}. This method has been extended to modern
spectrometers~\cite{violini_nima}. However, precise calculations using
analytical methods are complicated in the case of spectrometers with
multiple optical components, advanced neutron guide shapes and multiple
choppers. In this case, Monte-Carlo based numerical methods based can be
used to evaluate the resolution function \cite{mcstas, vickery_jpsj,
hahn_prb, granroth_wins}. Stimulated by preceding works, to estimate the
resolution function of the 4SEASONS spectrometer, which is operating at
J-PARC, we performed an McStas \cite{mcstas} simulation, based on the
model described in Ref.~\cite{4seasons}.

%The simulated resolution functions are displayed with the same
%instrument configuration as a typical single crystal experiment using
%the data analysis software suite for MLF, Utsusemi \cite{utsusemi}. In
%addition, we performed simple analytical calculations to validate the
%simulations.

4SEASONS is a time-of-flight direct geometry chopper spectrometer in the
Materials and Life Science Experimental Facility (MLF) at
J-PARC~\cite{4seasons}. It is installed at the BL01 beam port for
viewing the coupled moderator which is 18\,m upstream from the sample
position. Neutrons are transported by an elliptically converging guide
tube coated with supermirrors, whose cross-section at the exit is
43.4\,mm. The incident neutrons are monochromatized by a fast-rotating
Fermi chopper positioned 1.7\,m upstream from the sample position. In
addition, 4SEASONS has a T$_0$ chopper for suppressing fast neutrons and
two disk choppers for band definition. Neutrons scattered by a sample
are detected by the 19\,mm-diameter and 2.5\,m-long $^3$He position
sensitive detectors placed cylindrically at 2.5\,m from the sample
position. 4SEASONS was designed for measuring spin and lattice dynamics
in the 10$^0$--10$^2$\,meV energy range~\cite{seto_bba}. It originally
commissioned for a research project involving high-$T_c$ oxide
superconductors. Now, its use has expanded to studies of other
superconductors, magnetic materials, dielectrics, catalysts, and
thermoelectric materials.

\section{MONTE CARLO SIMULATION USING MCSTAS AND UTSUSEMI}

%\subsection{Outline of Evaluation of Resolution Function with McStas and
%  Utsusemi}

\subsection{Outline and Conditions of Simulation}

The model of 4SEASONS for the McStas simulation is basically same as
that used to simulate instrument performance~\cite{4seasons}, except for
the sample and detector, which are replaced by the
\texttt{TOFRes\_sample} and the \texttt{Res\_monitor} components,
respectively. The original simulation model well reproduces the
experimental neutron flux and energy resolution, as shown in
Ref.~\cite{4seasons}.  The T$_0$ chopper and the disk choppers were
excluded from the simulation model.  Considering that 2D magnetic
excitations observed in materials such as copper oxide superconductors
and iron-based superconductors account for most of the experiments
performed using 4SEASONS~\cite{sato_jpsj, ishii_natcom, horigane_scirep,
matsuura_prb}, in the present simulation, we assumed that the sample is
a 2D system, and its 2D plane is perpendicular to $\mathbf{k}_i$. The
sample is rod-shaped, and its diameter and height are denoted by $w_s$
and $h_s$, respectively. We studied samples of two sizes: one is
$w_s\!=\!h_s\!=\!20\,\mathrm{mm}$ and the other is
$w_s\!=\!h_s\!=\!40\,\mathrm{mm}$. The former was the size assumed in
designing the instrument~\cite{4seasons}, and the latter is the typical
size of the largest samples measured using the instrument. The assumed
crystal structure is orthorhombic with lattice parameters
$a\!\approx\!b=\!5.34\,\mathrm{\AA}$ and
$c\!=\!13.24\,\mathrm{\AA}$. The 2D plane is defined by the $a$ and the
$b$ axes. These values are typical for transition metal oxides with
so-called K$_2$NiF$_4$ structures. The $[001]$ axis is parallel to the
incident beam, and the $[100]$ axis is on the horizontal plane and
perpendicular to the incident beam. This configuration is similar to
those often employed in measurements of 2D systems~\cite{sato_jpsj,
ishii_natcom, horigane_scirep, matsuura_prb}. The incident energy
($E_i$) is 71\,meV, and the rotating frequency of the Fermi chopper
($f$) is 250\,Hz.

%This conversion is performed in the preprocessing part of the simulation
%code, i.e. \texttt{INITIALIZE} section in \texttt{.instr} file.

Simulations of the resolution function were performed using the
\texttt{TOFRes\_sample} and the \texttt{Res\_monitor} components in
McStas. The \texttt{TOFRes\_sample} is a sample component for
computation of the resolution function. It scatters neutrons
isotropically within a specified angular range and selects the neutron
energy uniformly so that neutron arrival times at the detector lie
within a specified time bin. The \texttt{TOFRes\_sample} is used
together with the \texttt{Res\_monitor}. It is a monitor component that
records all scattering events. The output file of the
\texttt{Res\_monitor} is a list of ``$\mathbf{k}_i$ (wave vector of
incident neutrons), $\mathbf{k}_f$ (wave vector of scattered neutrons),
position, intensity'' of the neutrons that reach the detector at a time
within the specified time bin~\cite{mcstas_manual}. Therefore, we need
to specify the detector pixel position and the time corresponding to
$\mathbf{Q}$ (momentum transfer) and $E$ (energy transfer) at which the
resolution is estimated. The position of $\mathbf{Q}$ is specified by
the 2D momentum transfer $\mathbf{Q}_\mathrm{2D}\!=\!(H,K)$ in
reciprocal lattice units (r.\,l.\,u.). The component of $\mathbf{Q}$
parallel to $\mathbf{k}_i$ ($L$ in r.\,l.\,u.) and the scattering angle
($\Phi$) vary as functions of $E$ following momentum and energy
conservations. We convert specified $\mathbf{Q}_\mathrm{2D}$ and $E$ to
the corresponding detector position in real space and the neutron
arrival time at the detector. The detector is a cylinder with a diameter
of 19\,mm and a height of 25\,mm , which are the same dimensions as that
of one detector pixel in 4SEASONS. The time bin at the detector is
1\,$\mu$s.

%To obtain the intrinsic energy resolution, time bin at the detector
%should be small as possible. In the present simulation, we assumed it is
%1\,$\mu$s.

\begin{figure}
 \centerline{\includegraphics[width=0.75\textwidth]{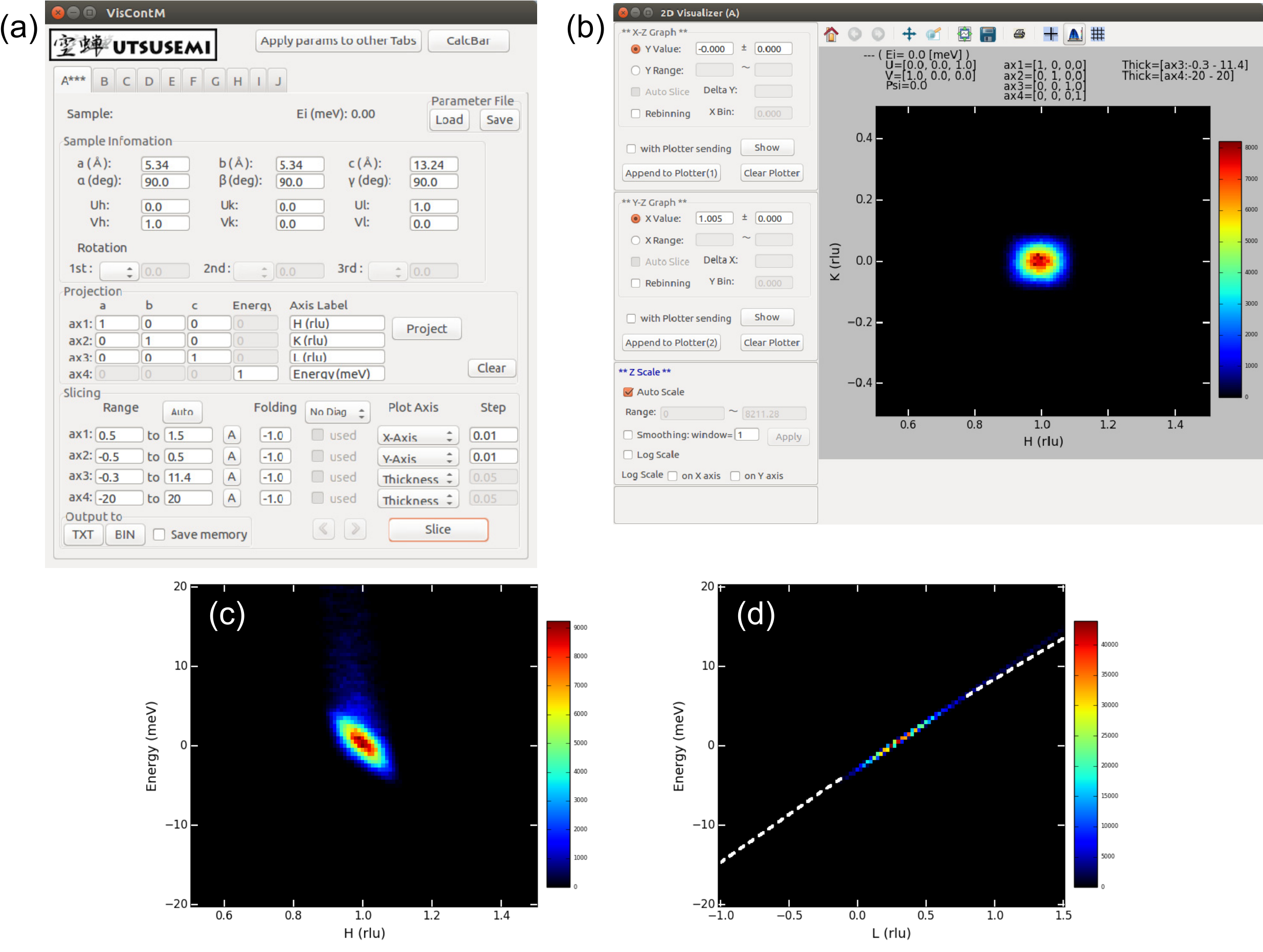}}
 \caption{(a) Graphical user interface of \texttt{VisContM} to visualize
 single crystal data. (b)--(d) Simulated resolution ellipsoid at
 $\mathbf{Q}_\mathrm{2D}\!=\!(1,0)$ ($a\!=\!5.34\,\mathrm{\AA}$) and
 $E=0\,\mathrm{meV}$ for $E_i\!=\!71\,\mathrm{meV}$,
 $f\!=\!250\,\mathrm{Hz}$, and $w_s\!=\!h_s\!=\!20\,\mathrm{mm}$,
 displayed on 2D maps as functions of (b) $H$ and $K$, (c) $H$ and $E$,
 and (d) $L$ and $E$. The dotted line in (d) shows a scan trajectory
 through $(1,0)$.}
 \label{utsusemi_panels}
\end{figure}

To visualize the simulated resolution function, we converted the
generated list of ``$\mathbf{k}_i$, $\mathbf{k}_f$, position, and
intensity'' to a list of ``$\mathbf{Q}$\,(\AA$^{-1}$), $E$\,(meV),
intensity, and error'', and visualized it by using the $Q$-$E$
visualizer in Utsusemi for single-crystal experiments,
\texttt{VisContM}~\cite{utsusemi}. Here $\mathbf{Q}$ and $E$ are given
by $\mathbf{Q} = \mathbf{k}_i - \mathbf{k}_f$ and $E = E_i - E_f =
\frac{\hbar^2}{2m_n}(k_i^2 - k_f^2)$, respectively, where $E_i$, $E_f$,
and $m_n$ are incident and scattered neutron energies, and neutron
mass. The $z$ axis is parallel to the incident beam, and the $x$ and $y$
axes are along the horizontal and vertical directions, respectively, in
a right-handed coordinate system.  By specifying crystallographic
information in \texttt{VisContM}, momentum transfer can be visualized in
r.\,l.\,u.\ in a manner similar to that in real experiments
[Fig.~\ref{utsusemi_panels}(a)]. The original \texttt{VisContM} handles
binary intensity data histogrammed as functions of detector pixel
positions and $E$, and averages counts in a bin upon slicing
data. However, it is not compatible with the list of resolution data
which is a text file storing randomly generated (not histogrammed)
$\mathbf{Q}$ and $E$ values of neutrons. Then, we added two new
functions to \texttt{VisContM}: one is a function to read a text file in
the form of [$\mathbf{Q}$\,(\AA$^{-1}$), $E$\,(meV), intensity, error]
and the other is a new slice mode that simply sums all counts in a bin.

\begin{figure}
 \centerline{\includegraphics[scale=0.55]{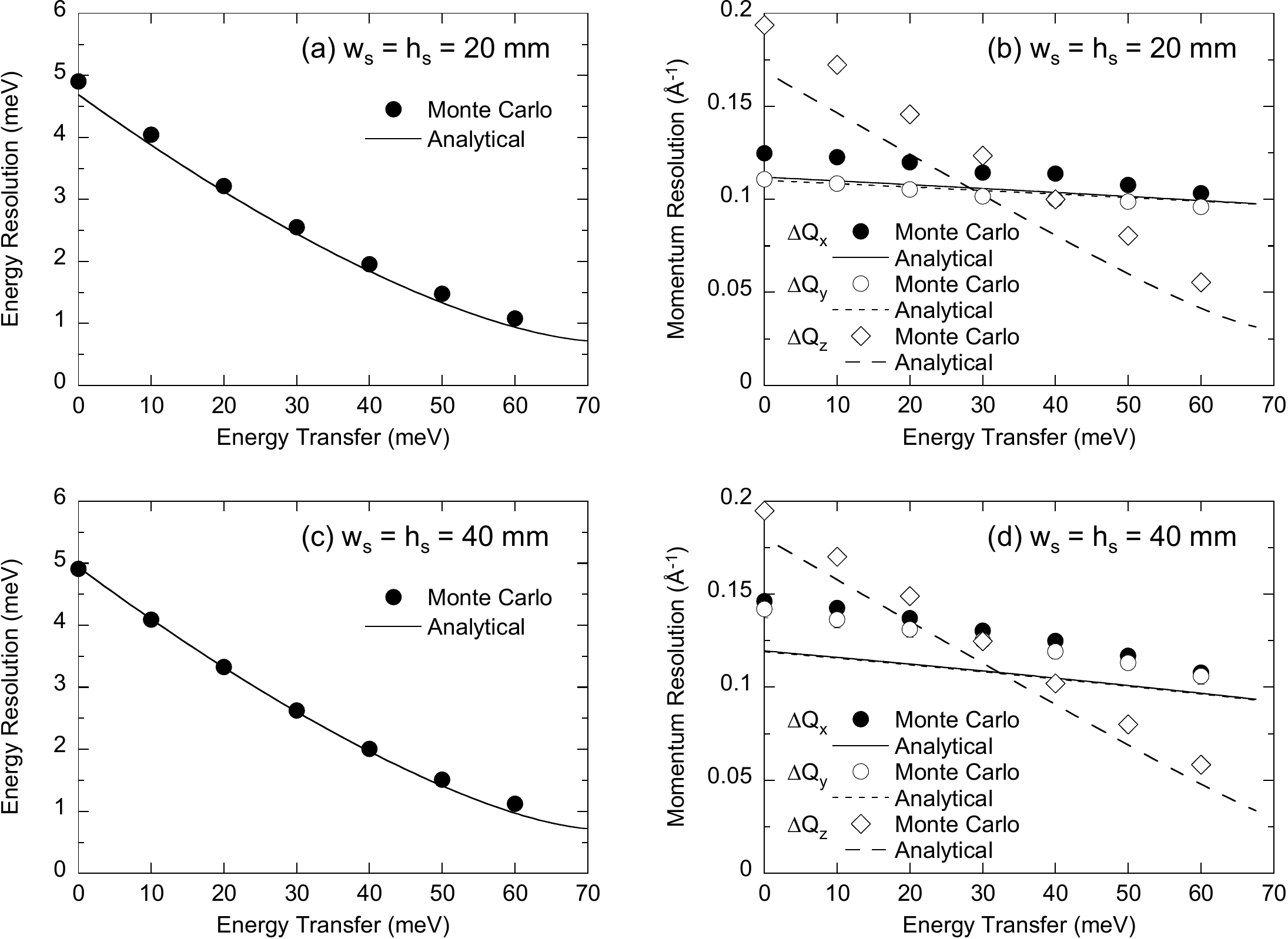}}
 \caption{Energy and momentum resolutions (FWHMs) for
 $E_i\!=\!71\,\mathrm{meV}$ and $f\!=\!250\,\mathrm{Hz}$ at
 $\mathbf{Q}_\mathrm{2D}\!=\!(1,0)$ ($a\!=\!5.34\,\mathrm{\AA}$) as
 functions of $E$. The energy resolution is evaluated for positive
 energy transfers (neutron energy loss side). (a) and (b) for
 $w_s\!=\!h_s=20\,\mathrm{mm}$. (c) and (d) for
 $w_s\!=\!h_s=40\,\mathrm{mm}$. In (a) and (c), the closed circles and
 the solid line show $\Delta E$ obtained by the Monte Carlo simulation
 and the analytical calculation, respectively. In (b) and (d), the
 closed circles, open circles, and open diamonds show the $\Delta Q_x$,
 $\Delta Q_y$, and $\Delta Q_z$ obtained by the Monte Carlo simulation,
 respectively. The solid, dotted, and broken lines show $\Delta Q_x$,
 $\Delta Q_y$, and $\Delta Q_z$ obtained by the analytical calculation,
 respectively.}
 \label{resolutions}
\end{figure}

\begin{figure}
 \centerline{\includegraphics[scale=0.50]{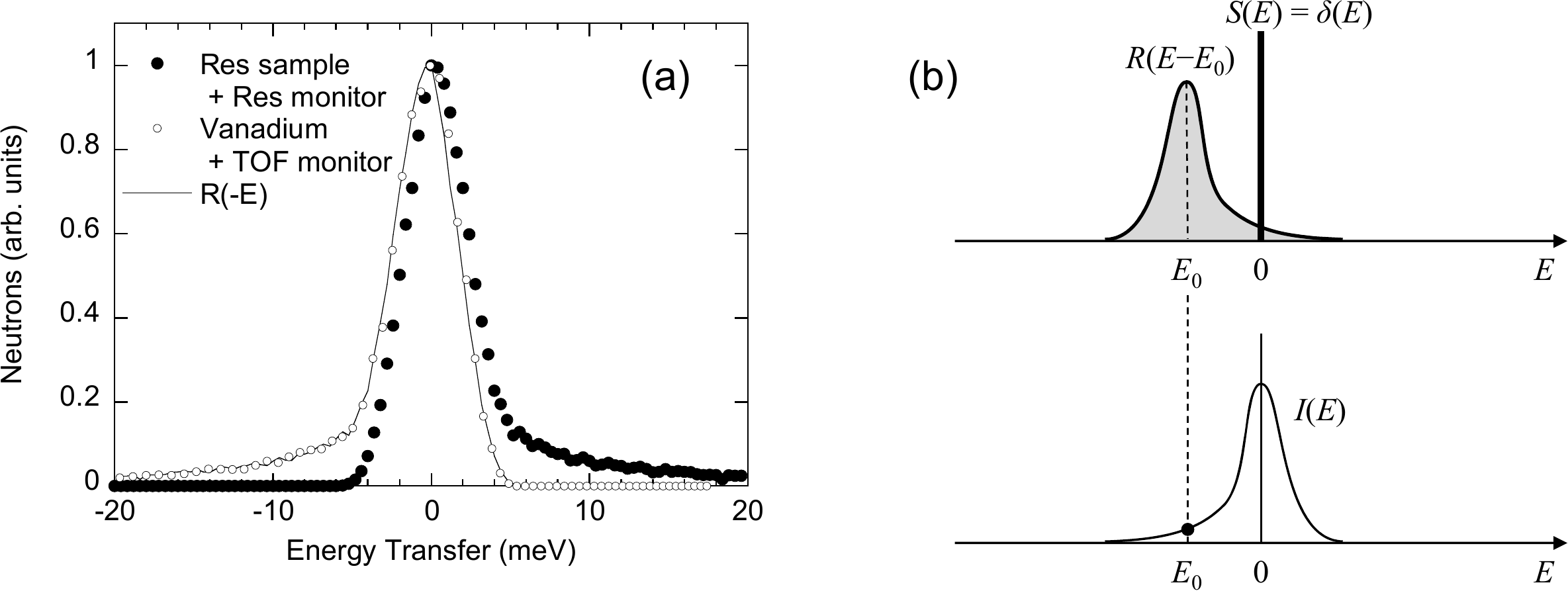}}
 \caption{(a) Comparison of Monte Carlo simulations of energy
 resolution for elastic scattering (closed circles) and energy
 spectrum for vanadium (open circles). For the simulation of vanadium,
 the \texttt{Incoherent} and \texttt{TOF\_monitor} components were used
 for the sample and detector, respectively, and the time spectrum is
 converted to the spectrum of $E$. For this simulation of the
 resolution, the ``square'' shape of the \texttt{Res\_monitor} component
 was chosen to be consistent with that of the \texttt{TOF\_monitor}
 component. Both data are normalized so that the peak heights
 are unity. The solid line is the inverted resolution function with
 respect to $E\!=\!0$. (b) Schematic drawings of the scattering function
 $S$, resolution function $R$, and observed spectrum $I$.}
 \label{tail}
\end{figure}

%The original \texttt{VisContM} handles binary histogramed
%intensity data as functions of detector pixel positions and $E$\,(meV)
%generated from a event-recorded raw data by histograming functions in
%Utsusemi, and projects it on the $\mathbf{Q}$\,(r.\,l.\,u.)  and
%$E$\,(meV) space using a specified crystallographic information. After
%these treatments, the projected 4D data can be sliced into a 2D map with
%given two axes to be visualized~\cite{utsusemi}.

%The latter is to handle the output of McStas, because
%it is not histogramed data but a list of randomly generated neutrons. As
%a result of the above customization of Utsusemi, the resolution
%ellipsoid can be visualized using the panel of \texttt{VisContM}
%similarly to real experiments [Fig.~\ref{utsusemi_panels}(a)].

%\subsection{Conditions}

\subsection{Results}

Figures~\ref{utsusemi_panels}(b)--(d) show a simulated resolution
function at $\mathbf{Q}_\mathrm{2D}\!=\!(1,0)$ and
$E\!=\!0\,\mathrm{meV}$ for $E_i\!=\!71\,\mathrm{meV}$,
$f\!=\!250\,\mathrm{Hz}$, and $w_s\!=\!h_s\!=\!20\,\mathrm{mm}$ when
$\mathbf{c} \parallel \mathbf{k}_i$ and $\mathbf{a}$ is on the
horizontal plane. To visualize the resolution function on 2D maps, we
integrated it over two axes in the four momentum and energy axes [axes
labeled ``Thickness'' in Fig.\,\ref{utsusemi_panels}(a)] and plotted it
as functions of the other two axes [axes labeled as ``X-axis'' and
``Y-axis'' in Fig.\,\ref{utsusemi_panels}(a)]. The resolution function
is generally expressed as an ellipsoid elongated and inclined with
respect to the $H$, $K$, $L$, and $E$
axes. Figure~\ref{utsusemi_panels}(d) shows it is especially elongated
along the $L$ or $E$ directions. This reflects the fact that the $L$
direction is parallel to the incident beam in the present simulation,
and therefore $L$ and $E$ are correlated strongly. In other words, the
resolution ellipsoid is elongated along the scan trajectory in the
$\mathbf{Q}$-$E$ space [see the dotted line in
Fig.~\ref{utsusemi_panels}(d)]. By contrast, when we see the resolution
ellipsoid on the $H$-$K$ plane, it is symmetric with respect to the $H$
and $K$ axes. This means that in a scattering configuration with
$\mathbf{k}_i$ and $\mathbf{k}_f$ in the horizontal plane, the
horizontal and vertical components of the $Q$ resolution are not
correlated.

To evaluate the simulated resolutions quantitatively, we projected the
4D resolution data onto one of the energy or momentum axes by
integrating them over the other three axes. Then, we fit the resulting
1D resolution profiles to Gaussians to obtain their full widths at half
maximum (FWHMs) as the resolution values. The closed circles in
Fig.~\ref{resolutions}(a) show the energy resolution ($\Delta E$) for
$w_s\!=\!h_s\!=\!20\,\mathrm{mm}$ as functions of $E$. The closed
circles, open circles, and open diamonds in Fig.~\ref{resolutions}(b)
show the momentum resolutions along the $H$, $K$, and $L$ axes,
respectively, for $w_s\!=\!h_s\!=\!20\,\mathrm{mm}$ as functions of
$E$. Given that in the present configuration the $H$, $K$, and $L$ axes
are parallel to the $x$, $y$, $z$ axes, respectively, we represent their
resolution components as $\Delta Q_x$, $\Delta Q_y$, and $\Delta
Q_z$. The symbols in Figs.~\ref{resolutions}(c) and \ref{resolutions}(d)
show the energy and the momentum resolutions for the case of
$w_s\!=\!h_s\!=\!40\,\mathrm{mm}$. As $E$ increases, both the energy and
the momentum resolutions decrease (become better). For the momentum
resolutions, $\Delta Q_z$ shows a faster decrease compared to $\Delta
Q_x$ and $\Delta Q_y$. The increase in sample size from
$w_s\!=\!h_s\!=\!20\,\mathrm{mm}$ to $w_s\!=\!h_s\!=\!40\,\mathrm{mm}$
increases $\Delta Q_x$ and $\Delta Q_y$, while it has little effect on
the energy resolution and $\Delta Q_z$.

In addition to the widths of the resolution functions, Monte Carlo
simulation is also useful for evaluating asymmetric tail in energy
spectrum often observed at chopper spectrometers at a spallation
source. The tail comes from the pulse tail at the moderator, and it is
observed at the neutron energy gain side owing to the pin-hole camera
effect of the chopper in the time-length space~\cite{arai_icans}. It is
more visible on a spectrometer for viewing a coupled moderator such as
4SEASONS. The closed circles in Fig.~\ref{tail}(a) show the resolution
function for elastic scattering integrated over all momentum axes, which
clearly shows the tail on the neutron energy loss (positive $E$) side in
contrast to the experiments [Tails can be seen similarly in
Figs.~\ref{utsusemi_panels}(c) and \ref{utsusemi_panels}(d).]. Here, we
should recall that the observed intensity at a particular momentum
transfer $\mathbf{Q}_0$ and $E_0$, $I(\mathbf{Q}_0,E_0)$ is a
convolution of the scattering function $S(\mathbf{Q},E)$ and the
resolution function $R(\mathbf{Q}-\mathbf{Q}_0,E-E_0)$, which is
expressed as $I(\mathbf{Q}_0,E_0) = \int\!
R(\mathbf{Q}-\mathbf{Q}_0,E-E_0)S(\mathbf{Q},E)\,d\mathbf{Q}dE$. If
$S(\mathbf{Q},E) = \delta(E)$, as in the case of elastic incoherent
scattering, $I(E_0) = R(-E_0)$. Then, the tail appears on the opposite
side, that is, the negative $E$ side, in experiments as shown
schematically in Fig.~\ref{tail}(b). The solid line in Fig.~\ref{tail}
shows the same resolution function inverted with respect to $E=0$, which
coincides with the simulated scattering profile of vanadium (open
circles).

\section{COMPARISON WITH ANALYTICAL CALCULATION}

To validate the resolutions evaluated by the Monte Carlo simulation, we
performed analytical calculations of the energy and momentum
resolution. In general, analytical calculation of the resolution
ellipsoid in the 4D $\mathbf{Q}$-$E$ space is complicated. Here,
instead, we consider a less general case of scattering in the horizontal
plane, for which a streamlined calculation is used to obtain the energy
and momentum resolution (FWHM). FWHM is proportional to its standard
deviation if the distribution is a Gaussian.

%Assuming the distribution of each resolution component has a Gaussian
%distribution, we calculated the resolution function as square-root of
%sums of squares of resolution components.

%we performed simple and intuitive calculations of energy resolution of
%incoherent scattering and momentum resolutions along principal axes,
%assuming the scattering plane is confined on the horizontal plane. We
%represent the resolutions as well as the deviation of each resolution
%element with respect to its nominal value as their FWHMs.

\subsection{Energy Resolution}

%Since the energy resolution of 4SEASONS has been already
%established~\cite{4seasons,iida_jpsconf}, we here briefly summarize its
%expression. The energy resolution of a Fermi chopper spectrometer is
%given by~\cite{windsor,arcs,cncs}

We note here that the simplified expression used
elsewhere~\cite{4seasons,iida_jpsconf,windsor,arcs,cncs} to calculate
the energy resolution (FWHM) of the 4SEASONS spectrometer:
\begin{equation}
 \Delta E = 2E_i\left(
                 \left\{
                  \frac{\Delta t_c}{t_c}
                  \left[1+\frac{L_1}{L_2}
                       \left(1-\frac{E}{E_i}\right)^{\frac{3}{2}}
                  \right]
                 \right\}^2
		 +
		 \left\{
		  \frac{\Delta t_m}{t_c}
		  \left[1+\frac{L_3}{L_2}
		       \left(1-\frac{E}{E_i}\right)^{\frac{3}{2}}
		  \right]
		 \right\}^2
		 +
		 \left[
		  \frac{\Delta L_2}{L_2}
		  \left(1-\frac{E}{E_i}\right)\right]^2
	 \right)^{1/2}.
 \label{Ereso_eq}
\end{equation}
$E_i$ and $E$ are the incident energy and energy transfer,
respectively. $L_1$, $L_2$, and $L_3$ are the moderator-to-sample,
sample-to-detector, and chopper-to-sample distances, respectively. For
4SEASONS, $L_1\!=\!18\,\mathrm{m}$, $L_2\!=\!2.5\,\mathrm{m}$, and
$L_3\!=\!1.7\,\mathrm{m}$. $\Delta t_c$ and $\Delta t_m$ are the opening
time of the Fermi chopper and the pulse width at the moderator,
respectively. $\Delta t_c$ is effectively larger than its intrinsic
value defined by chopper geometry, $\Delta t_\mathrm{c0} =
d/(2{\pi}Df)$, because of angular divergence of the incident beam as
$\Delta t_c = \Delta t_\mathrm{c0}\cdot p(\Delta\Phi_i^\mathrm{max})$.
$D$ and $d$ are the diameter of the chopper rotor and the width of each
of the slits, respectively, and $d/D\!=\!0.02$ for 4SEASONS. $\Delta
\Phi_i^\mathrm{max}$ is the maximum angular divergence of the incident
beam, and the function $p$ is described in Ref.~\cite{windsor}.
$\Delta\Phi_i^\mathrm{max}$ has neutron energy dependence originating
from neutron reflections by the supermirrors of the guide tube. It was
estimated using the relationship between neutron wavelength and
supermirror critical angle as described in
Ref.~\cite{ikeuchi_jpsj}. $\Delta t_m$ varies with
$E_i$~\cite{ikeda-carpenter}, and we used the numerical values of
$\Delta t_m$ from Ref.~\cite{pulsedata}.  $\Delta L_2$ is the
uncertainty of $L_2$ resulting from the sample and the detector
sizes. In general, $\Delta L_2$ depends on the sample shape, and the
shape dependence should be considered seriously for high-resolution
time-of-flight spectrometers~\cite{zorn_nima}. Here, however, to
consider the resolution of the middle-resolution spectrometer, we
assumed $\Delta L_2 = [(\pi w_s/4)^2 + (\pi w_d/4)^2]^{1/2}$, where
$w_s$ and $w_d$ are diameters of the sample and detector,
respectively. $w_d$ is 19\,mm for 4SEASONS. $\pi w_s/4$ ($\pi w_d/4$) is
the average transmission length of a cylinder for a beam perpendicular
to the cylinder. $t_c$ is the time at which neutrons with energy $E_i$
reach the Fermi chopper.

The solid lines in Figs.~\ref{resolutions}(a) and \ref{resolutions}(c)
show the $\Delta E$ calculated using Eq.~(\ref{Ereso_eq}) for
$E_i\!=\!71\,\mathrm{meV}$ and $f\!=\!250\,\mathrm{Hz}$ in the cases of
$w_s=20\,\mathrm{mm}$ and $w_s=40\,\mathrm{mm}$, respectively. They show
excellent agreement with the results of the Monte Carlo simulations
(closed circles). $\Delta E$ is dominated by the contribution of the
first term in Eq.~(\ref{Ereso_eq}), and the other two terms have smaller
contributions~\cite{iida_jpsconf}. Therefore, the increase in sample
size, which increases $\Delta L_2$, has little effect on the energy
resolution, as shown in Figs.~\ref{resolutions}(a) and
\ref{resolutions}(c).

\subsection{Momentum Resolution}

\begin{figure}
  \centerline{\includegraphics[scale=0.48]{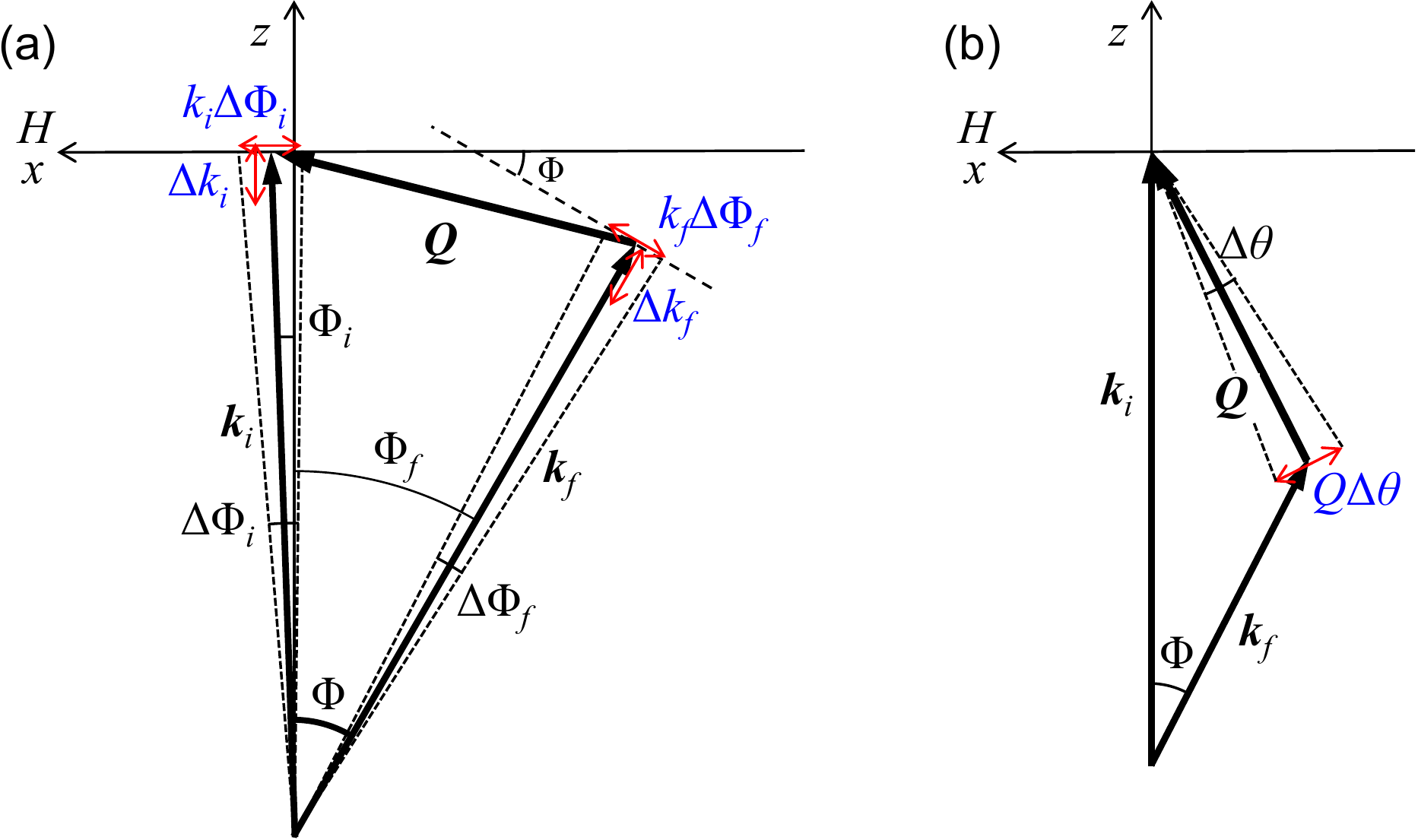}}
  \caption{(a) Scattering diagram considered in present
  calculations. (b) Scattering diagram showing effect of sample
  mosaicity.}
  \label{scattering_diagram}
\end{figure}

Here, we concentrate on deriving the resolution components along the
principal axes to compare them with the simulation results shown in
Fig.~\ref{resolutions}(b). First, the horizontal components of the
momentum resolution are calculated. The $x$ and $z$ components of
$\mathbf{Q}$ are defined as
\begin{equation}
 \begin{array}{rcl}
  Q_x &=& k_i \sin(\Phi_i) - k_f \sin(\Phi_f), \\[\jot]
  Q_z &=& k_i \cos(\Phi_i) - k_f \cos(\Phi_f),
 \end{array} 
\label{QH}
\end{equation}
where $\Phi_i$ and $\Phi_f$ are the angles of $\mathbf{k}_i$ and
$\mathbf{k}_f$ relative to the primary beam centerline, respectively
[Fig.~\ref{scattering_diagram}(a)], and nominally $\Phi_i = 0$. By
differentiating Eq.~(\ref{QH}) with respect to each parameter, we obtain
the deviation of $Q_x$ from its nominal value as follows:
\begin{eqnarray}
 \Delta Q_x
  &=& \left[
      \left(\left.\frac{\partial Q_x}{\partial\Phi_i}\right|_{\Phi_i=0}
       \!\!\Delta\Phi_{i}\right)^2
    + \left(\left.\frac{\partial Q_x}{\partial\Phi_f}\right|_{\Phi_f=\Phi}
       \!\!\Delta\Phi_{f}\right)^2
    + \left(\left.\frac{\partial Q_x}{\partial k_i}\right|_{\Phi_i=0}
       \!\!\Delta k_i\right)^2
    + \left(\left.\frac{\partial Q_x}{\partial k_f}\right|_{\Phi_f=\Phi}
       \!\!\Delta k_f\right)^2 \right]^{1/2} \nonumber \\
  &=& \left[
      (k_i \Delta\Phi_{i})^2
    + (k_f \Delta\Phi_{f} \cos\Phi)^2
    + (\Delta k_f \sin\Phi)^2 \right]^{1/2} ,
\label{DQx}
\end{eqnarray}
where $\Delta\Phi_{i}$ ($\Delta\Phi_{f}$) is the angular divergence of
the incident (scattered) beam on the horizontal plane, and $\Delta k_i$
($\Delta k_f$) is the uncertainty in $k_i$ ($k_f$)
[Fig.~\ref{scattering_diagram}(a)]. Similarly, for $Q_z$, we obtain
\begin{equation}
 \Delta Q_z = \left[(k_f \Delta\Phi_{f} \sin\Phi)^2
                  + (\Delta k_i)^2
                  + (\cos\Phi \Delta k_f)^2 \right]^{1/2} .
\label{DQz}
\end{equation}
The contributions of each of the terms in Eqs.~(\ref{DQx}) and (\ref{DQz}) can
be understood geometrically from the scattering diagram in
Fig.~\ref{scattering_diagram}(a).

$\Delta k_i$ and $\Delta k_f$ are calculated by differentiating $k_i =
\frac{m_n}{\hbar}(L_1 - L_3)/t_c$ and $k_f = \frac{m_n}{\hbar}L_2/(t_d -
t_s)$ with respect to $t_c$, $t_s$, and $L_2$, where $t_s$ ($t_d$) is
the arrival time of neutrons at the sample (detector). Given that the
uncertainties in $t_c$ and $t_s$ are $[(\Delta t_c)^2 + (\Delta
t_m)^2]^{1/2}$ and $\left\{[\Delta t_c L_1/(L_1-L_3)]^2 + [\Delta t_m
L_3/(L_1-L_3)]^2\right\}^{1/2}$, respectively,
\begin{equation}
 \Delta k_i
  = k_i \frac{\left[ (\Delta t_c)^2 + (\Delta t_m)^2 \right]^{1/2}}{t_c},
 \label{Dki}
\end{equation}
\begin{equation}
 \Delta k_f
  = k_f \left[
        \left(\frac{L_1}{L_1-L_3}\frac{\Delta t_c}{t_d-t_s}\right)^2
      + \left(\frac{L_3}{L_1-L_3}\frac{\Delta t_m}{t_d-t_s}\right)^2
      + \left(\frac{\Delta L_2}{L_2}\right)^2
	\right]^{1/2} .
 \label{Dkf}
\end{equation}
By substituting Eqs.~(\ref{Dki}) and (\ref{Dkf}) into Eqs.~(\ref{DQx})
and (\ref{DQz}), we obtain the expressions of $\Delta Q_x$ and $\Delta
Q_z$ (they fall into the same expressions as written in Ref.~\cite{cncs}
except for the $\Delta\Phi_{i}$ term in Eq.~(\ref{DQx}).). As for the
angular divergences, $\Delta\Phi_{f}$ is determined by the sample size
and the detector pixel size. By neglecting the shapes of the sample and
detector, $\Delta\Phi_{f} = (w_s + w_d)/(2L_2)$. By contrast, it is
difficult to describe $\Delta\Phi_i$ by using a simple function. We
confirmed by simulations that although the maximum divergence of the
incident beam can be well reproduced by the supermirror critical angle,
which was used to calculate the $E$ resolution, the profile shape and
the FWHM of the angular divergence deviate from the Gaussian and depends
strongly on incident energy and sample size. Then, we fitted the
simulated angular divergence of incident neutrons hitting the sample to
a Gaussian, and used its FWHM as $\Delta\Phi_{i}$.

%$\Delta\Phi_{i}$ was estimated from the neutron-wavelength dependence of
%the supermirror critical angle in the same manner as used in the
%estimation of the energy resolution.

The vertical component of the momentum resolution $\Delta Q_y$ was
obtained using the divergences of the incident and the scattered beams:
\begin{equation}
 \Delta Q_y = \left[(k_i\Delta\Phi_{i,v})^2 + (k_f\Delta\Phi_{f,v})^2\right]^{1/2}
\label{DQy}
\end{equation}
where $\Delta\Phi_{i,v}$ and $\Delta\Phi_{f,v}$ are angular divergences
of the incident and the scattered beams along the vertical direction,
respectively. Given that the cross-section of the neutron guide tube of
4SEASONS has a square shape, $\Delta\Phi_{i,v}$ is equal to
$\Delta\Phi_{i}$ in Eq.~(\ref{DQx}). As for $\Delta\Phi_{f,v}$, by using
the sample height $h_s$ and detector pixel height $h_d$,
$\Delta\Phi_{f,v} = (h_s + h_d)/(2L_2)$. $h_d\!=\!25\,\mathrm{mm}$ for
4SEASONS.

The solid, dotted, and broken lines in Figs.~\ref{resolutions}(b) and
\ref{resolutions}(d) show the calculated values of $\Delta Q_x$, $\Delta
Q_y$, and $\Delta Q_z$, respectively, at
$\mathbf{Q}_\mathrm{2D}\!=\!(1,0)$ ($a\!=\!5.34\,\mathrm{\AA}$) for
$E_i\!=\!71\,\mathrm{meV}$ and $f\!=\!250\,\mathrm{Hz}$. They show
reasonably good agreement with the simulations. The increase in sample
size leads to an increase in $\Delta \Phi_f$ ($\Delta \Phi_{f,v}$) and
$\Delta L_2$, resulting in an increase in the momentum resolution
values. In the present calculations, the three terms in $\Delta Q_x$
have comparable contributions at $E\!=\!0$. As $E$ increases, the
$\Delta\Phi_f$ and $\Delta k_f$ terms decrease because $k_f$ decreases
and $\Phi$ increases, while the $\Delta \Phi_i$ term is independent of
$E$. Therefore, the incident beam divergence is critical for the
momentum resolution in such experiments, where $Q_x$ is of particular
interest, which is often the case in experiments involving 1D or 2D
systems. By contrast, in $\Delta Q_z$, the $\Delta k_f$ term is
dominant, while the $\Delta \Phi_f$ term and the $\Delta k_i$ term are
small and independent of $E$ (Note that $k_f \sin\Phi = Q_x$ is fixed in
the present calculation.). The rapid decrease in $\Delta Q_z$ as a
function of $E$ follows the behavior of $\Delta E$ in
Fig.~\ref{resolutions}(a) and \ref{resolutions}(c). This comes from the
fact that $\Delta k_f$ is strongly correlated with the energy
resolution. As for $\Delta Q_y$, it is determined only by the angular
beam divergences, which are independent of $E$. $\Delta Q_y$ depends
weakly on $E$ through the dependence of $k_f$ on $E$.

\begin{figure}
 \centerline{\includegraphics[scale=0.55]{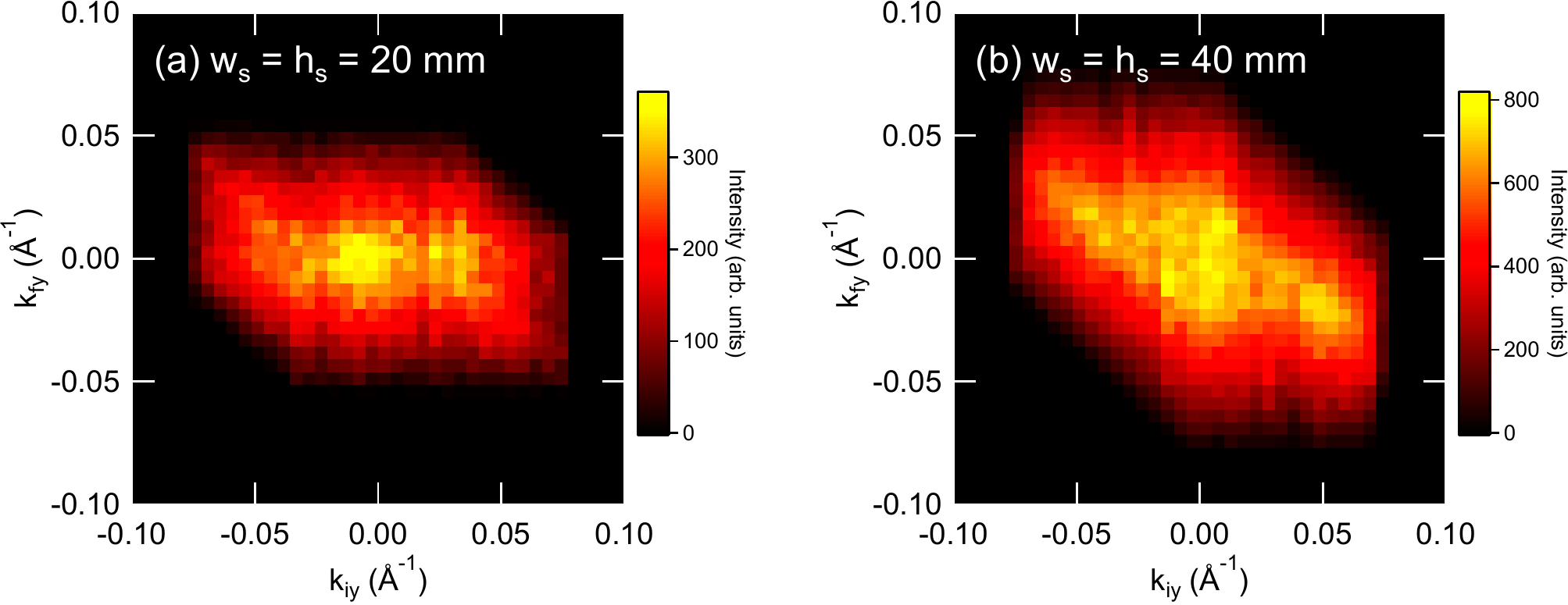}}
 \caption{Correlations between $k_{iy}$ and $k_{fy}$ for (a)
 $w_s\!=\!h_s\!=\!20\,\mathrm{mm}$ and (b)
 $w_s\!=\!h_s\!=\!40\,\mathrm{mm}$ obtained the Monte Carlo simulation
 and shown as neutron distribution on the $k_{iy}$-$k_{fy}$ planes.}
 \label{kikfcorrelations}
\end{figure}

The difference between the simulations and analytical calculations
probably comes from the rather simple modeling in the analytical
case. As shown in Fig.~\ref{utsusemi_panels}, $\Delta Q_x$ and $\Delta
Q_z$ are correlated strongly with the $E$ component of the resolution
function. Neglecting these correlations may contribute to the
discrepancy. Moreover, because $\Delta Q_z$ is dominated by the $\Delta
k_f$ term, it can be affected by the asymmetric time structure of the
neutron beam due to the moderator pulse tail. By contrast, in
Fig.~\ref{resolutions}(b), $\Delta Q_y$ shows good agreement with the
simulation results, possibly because $\Delta Q_y$ is simply expressed in
terms of the angular divergence and the correlations with the other
resolution components are small. However, as the sample size increases
to $w_s\!=\!h_s\!=\!40\,\mathrm{mm}$, even the simulation and analytical
results of $\Delta Q_y$ show a significant discrepancy. Eq.~(\ref{DQy})
for $\Delta Q_y$ is a square-root of the sum of the squares of the
distributions of $k_{iy}$ ($\Delta k_{iy}$) and $k_{fy}$ ($\Delta
k_{fy}$). Therefore, we assumes that $\Delta k_{iy}$ and $\Delta k_{fy}$
are not correlated. To check whether this assumption holds in the
simulations, in Fig.~\ref{kikfcorrelations}, we plotted the simulated
neutron distributions as functions of $k_{iy}$ and $k_{fy}$ for the two
sample sizes. We found that a considerable correlation exists between
$\Delta k_{iy}$ and $\Delta k_{fy}$, and there are fewer neutrons with
$k_{iy} \sim k_{fy}$ in the $k_{iy}$-$k_{fy}$ map. Then, in the
distribution of $Q_y = k_{iy} - k_{fy}$, the contribution of $Q_y \sim
0$ should be suppressed, while that of $|Q_y| > 0$ should be enhanced,
resulting in an effective increase in FWHM. This correlation comes from
the fact that few neutrons that hit the sample upward (downward) and are
scattered by the sample upward (downward) again can reach the detector
pixel. As the sample size increases, the number of such unreached
neutrons increases and this geometrical effect becomes significant in
the momentum resolution. A similar geometrical correlation also exists
between $k_{ix}$ and $k_{fx}$, and it should become one of the causes of
the discrepancy in $\Delta Q_x$ between the simulated and the analytical
results.

%Furthermore, the good agreement for $\Delta Q_y$ shows the Gaussian
%approximation of the non-Gaussian angular divergence of the incident
%beam gives a good estimation of the resolution function.

%In particular, there may be ambiguities in approximating the incident
%beam divergence and time spectra which are not simple Gaussians due to
%effects of the supermirrors and the moderator pulse tail. Nevertheless,

\section{SUMMARY AND REMAINING ISSUES}

We performed Monte Carlo simulations of the resolution function for the
4SEASONS spectrometer by using McStas. Visualization of the obtained 4D
resolution functions by using Utsusemi, the data analysis suite for MLF,
facilitates intuitive understanding of the resolution
functions. Furthermore, we performed analytical calculations of the
energy and momentum resolutions. Although they were performed only for a
simple model, reasonably good agreements with the simulation results not
only validate the simulations, but also prove that such a simple
calculation is quite useful for obtaining a rough estimate of the
resolution.  We plan to develop the simulation toward more general cases
of 3D systems and enhance compatibility with the data analysis software.

Finally, we mention aspects that are important for resolution analysis
of experimental data but were not considered in the present study. In
this study, we performed the simulation with the smallest time bin at
the detector as possible to obtain the intrinsic energy resolution. In
real experiments, however, data is often cut with a finite width of $E$
(roughly, 1--10\,meV). To consider the width of the energy cut, we must
use a finite time bin at the detector that corresponds to the width of
the energy transfer.  Another important aspect is sample mosaicity. The
resolution obtained by McStas is the intrinsic resolution determined by
the geometrical arrangements of the instrument components. In this
sense, the effect of sample mosaicity is effectively included in the
dynamic structure factor $S(\mathbf{Q},E)$ of the sample. However, it is
often useful to combine the effects of sample mosaicity with the
resolution to deduce the \emph{true} $S(\mathbf{Q},E)$ representing
physical properties~\cite{shirane}. However, it seems that the
\texttt{Res\_sample} and the \texttt{TOFRes\_monitor} components in
McStas cannot treat sample mosaicity directly, and therefore, some
modifications to the simulation codes are required. It would be useful
to consider the simple case that sample mosaicity is represented as a
rotation of $\mathbf{Q}$ by $\Delta\theta$
[Fig.~\ref{scattering_diagram}(b)]. Then, the resolution perpendicular
to $\mathbf{Q}$ is most affected by the sample mosaicity. For example,
when we are interested only in $Q_x$, as is the case in measurements of
1D or 2D systems, the term originating from the sample mosaicity $\Delta
Q_\mathrm{mos}$ is $Q\Delta\theta\cos[\arcsin(Q_x/Q)]$. It has little
effect on $\Delta Q_x$ when $E$ is sufficiently small, but it should be
considered as $E$ increases.

\section{ACKNOWLEDGMENTS}

We thank K. Nakajima, M. Nakamura, G. E. Granroth, P. Willendrup and
E. Farhi for valuable discussions and useful advices. R. K. and
Y. I. were supported by JSPS KAKENHI Grant Number 15K04742.

% References

\end{document}